\documentclass[twocolumn]{aastex62}

\usepackage{amsmath}
\usepackage{bm}
\usepackage{stmaryrd}
\usepackage{mathtools}
\usepackage{xcolor}
\usepackage{soul}

\begin{document}

\title{An Analytical Model of the Kelvin-Helmholtz Instability of Transverse Coronal Loop Oscillations}

\correspondingauthor{Mihai Barbulescu}
\email{mbarbulescu1@sheffield.ac.uk}

\author[0000-0001-9569-8306]{Mihai Barbulescu}
\affiliation{Solar Physics and Space Plasma Research Centre, School of Mathematics and Statistics, University of Sheffield, Hicks Building, Hounsfield Road, Sheffield, S3 7RH, UK}

\author[0000-0003-2324-8466]{Michael S. Ruderman}
\affiliation{Solar Physics and Space Plasma Research Centre, School of Mathematics and Statistics, University of Sheffield, Hicks Building, Hounsfield Road, Sheffield, S3 7RH, UK}
\affiliation{Space Research Institute (IKI), Russian Academy of Sciences, 117997 Moscow, Russia}

\author[0000-0001-9628-4113]{Tom Van Doorsselaere}
\affiliation{Centre for mathematical Plasma Astrophysics, Department of Mathematics, KU Leuven, Celestijnenlaan 200B, bus 2400, 3001 Leuven, Belgium}

\author[0000-0003-3439-4127]{Robert Erd\'elyi}
\affiliation{Solar Physics and Space Plasma Research Centre, School of Mathematics and Statistics, University of Sheffield, Hicks Building, Hounsfield Road, Sheffield, S3 7RH, UK}
\affiliation{Department of Astronomy, E\"otv\"os Lor\'and University, Budapest, P\'azm\'any P. s\'et\'any 1/A, H-1117, Hungary}

\begin{abstract}

Recent numerical simulations have demonstrated that transverse coronal loop oscillations are susceptible to the Kelvin-Helmholtz (KH) instability due to the counter-streaming motions at the loop boundary.
We present the first analytical model of this phenomenon.
The region at the loop boundary where the shearing motions are greatest is treated as a straight interface separating time-periodic counter-streaming flows.
In order to consider a twisted tube, the magnetic field at one side of the interface is inclined.
We show that the evolution of the displacement at the interface is governed by Mathieu's equation and we use this equation to study the stability of the interface.
We prove that the interface is always unstable, and that, under certain conditions, the magnetic shear may reduce the instability growth rate.
The result, that the magnetic shear cannot stabilise the interface, explains the numerically found fact that the magnetic twist does not prevent the onset of the KH instability at the boundary of an oscillating magnetic tube.
We also introduce the notion of the loop $\sigma$-stability.
We say that a transversally oscillating loop is $\sigma$-stable if the KH instability growth time is larger than the damping time of the kink oscillation.
We show that even relatively weakly twisted loops are $\sigma$-stable.

\end{abstract}


%=============================================================
\section{Introduction}
%=============================================================

Transverse oscillations of coronal loops have been a subject of extensive study since their original observation on 14 July 1998 by the \textit{Transition Region and Coronal Explorer} (TRACE) \citep{Aschwanden1999, Nakariakov1999}. For a review of the theory of these oscillations see \cite{Ruderman2009}.

In particular, the damping mechanism of transverse loop oscillations has received much attention \citep[e.g.][]{Ruderman2002, Goossens2002, TVD2004, Dymova2006, Williamson2014}, with the caveat that many studies have relied on the assumption that the oscillations are in the linear regime.
The nonlinear damping of transverse coronal loop oscillations has also been studied, both analytically \citep{Ruderman2010, Ruderman2014, Ruderman2017}, as well as numerically \citep[e.g.][]{Terradas2004, Magyar2016a}.
The numerical studies revealed important effects, such as that of the ponderomotive force, and the presence of the Kelvin-Helmholtz instability (KHI) at the loop boundaries.
More recently, \cite{Goddard2016} carried out a statistical study of observations of the damping of coronal loop kink oscillations.

\begin{figure*}[t]
\centering
\includegraphics[width=0.9\textwidth]{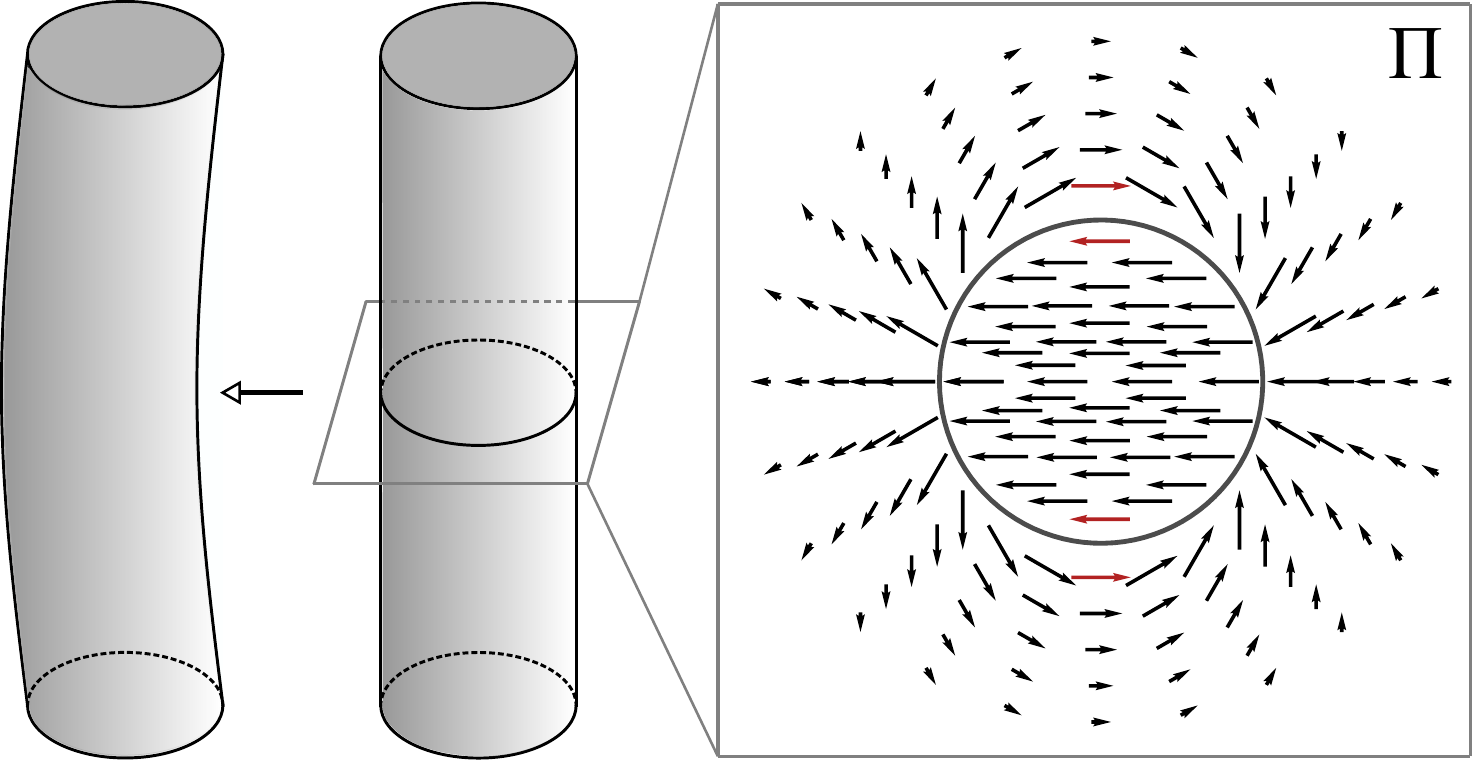}
\caption{Sketch of a straight magnetic flux tube with stationary footpoints undergoing transverse (kink) motion. The panel on the right represents the velocity field in a cross-section of the tube, at half the length of the tube. The greatest shearing occurs between the vectors coloured in red.}
\label{fig:tube1}
\end{figure*}

\cite{Terradas2008} suggested that a kink oscillation may render a flux tube unstable due to the shear motions at the boundaries.
The authors found that, for a smooth transition layer, the instability developed rapidly where the difference between the internal and external flow amplitudes was the greatest.
However, increasing the thickness of the transitional layer significantly decreased the growth rate of the instability.
It is worth noting that the KHI in smooth transition layers via other mechanisms (e.g. phase mixing, resonant absorption) had also received attention previously \citep[see, for example,][]{Heyvaerts1983, Ofman1994, Poedts1997}.
For a recent review on modelling the KHI see, e.g. \cite{Zhelyazkov2015}.

The topic of the transverse wave induced Kelvin-Helmholtz (TWIKH) instability was subsequently investigated by \cite{Antolin2014}, who suggested that this phenomenon may be responsible for the fine strand-like structure observed in some coronal loops. {In their numerical modelling these authors found that this structure is formed near the loop boundary even when the oscillation amplitude is very small, about 3~km/s. This result implies that the TWIKH instability develops even for very small oscillation amplitudes.} The TWIKH instability has since been studied by \cite{Antolin2016, Magyar2016a, Magyar2016b, Antolin2017, Karampelas2017, Howson2017a, Howson2017b, Karampelas2018}, who considered various aspects of the instability onset, growth rate and observational properties.

The configuration of the equilibrium magnetic field is an important aspect of TWIKH instabilities. It was suggested by \cite{Terradas2008} that a twisted magnetic field may suppress the instability. The effect of twist on the stability of transverse loop transverse oscillations was studied numerically by \cite{Howson2017b} who investigated the energetics of the instability of a magnetically twisted coronal loop and found that its evolution is affected by the strength of the azimuthal component of the magnetic field.
The authors also found that, when magnetic twist is present, the KHI leads to greater Ohmic dissipation as a result of the production of larger currents. Furthermore, \cite{Terradas2018} studied the evolution of the instability and found that the  magnetic twist increases the instability growth time.

Numerical simulations have provided some insight into the development of the KHI, but have not thoroughly established what the conditions are needed for its onset. In this paper, we find these requirements analytically by modelling the boundary of the flux tube where the shearing is greatest as a single interface separating regions of different densities and magnetic fields, and performing a local stability analysis. We emulate the effect of the transverse oscillation by subjecting each region to temporally periodic counter-streaming flows.

Although this work is the first local analysis of the TWIKH instability with oscillating flows, the KHI in the presence of transverse shear and twisted magnetic fields has previously been studied by \cite{Soler2010} and \cite{Zaqarashvili2015}.
The aforementioned studies, however, consider steady flows in a cylindrical geometry, while this paper is concerned with the analysis of temporally periodic flows in a Cartesian geometry.

The paper is organised as follows: in Section~\ref{sec:goveq}, we introduce a Cartesian model of the boundary of a twisted flux tube, and derive the governing equation for the displacement.
The stability of the flow is analysed in Section~\ref{sec:stab}, followed by applications to transverse coronal loop oscillations in Section~\ref{sec:loop}.
Section~\ref{sec:sum} contains the summary of the obtained results and our conclusions.

%=============================================================
\section{The Governing Equation}
\label{sec:goveq}
%=============================================================

\begin{figure*}[t]
\gridline{
\fig{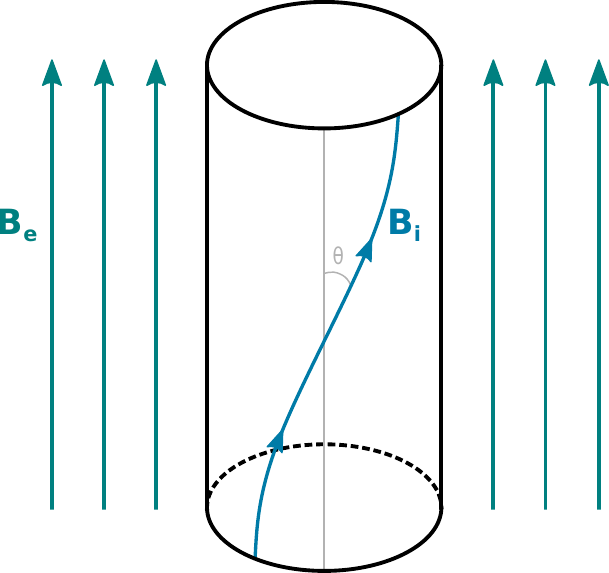}{0.9\columnwidth}
{(a) A twisted magnetic flux tube embedded in a straight magnetic field.}
\fig{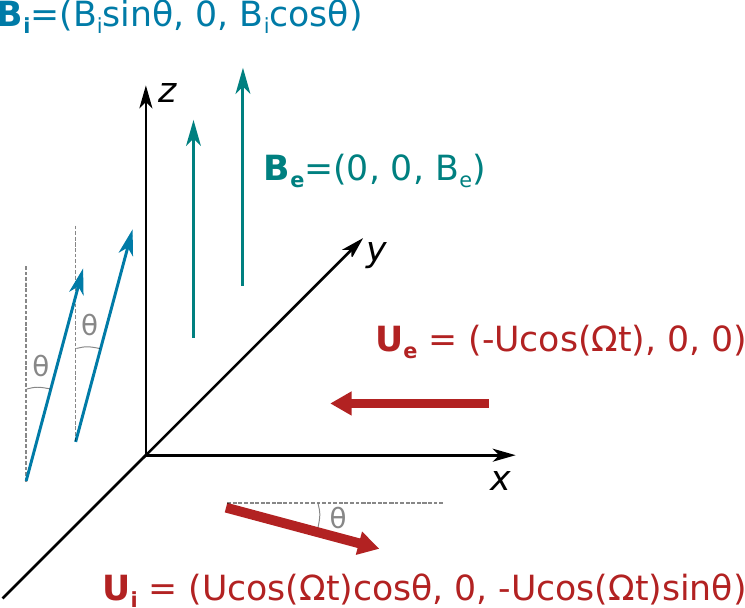}{0.9\columnwidth}{(b) The magnetic fields and flows at the interface.}
}
\caption{Sketch of a twisted magnetic tube, (a), and a diagram of the flows on each side of the boundary during transverse oscillation (b).}
\label{fig:tube_interface}
\end{figure*}

It is well established that a magnetic flux tube undergoing transverse oscillation is prone to the Kelvin-Helmholtz instability due to the shearing motions at the boundaries \citep{Terradas2008}.
Considering only the fundamental mode of oscillation, we wish to obtain the TWIKH instability criterion.
We start by considering a magnetically twisted flux tube of length $L$.
{For mathematical simplicity, we consider the boundary of the tube to be a  tangential interface, meaning there is no smooth boundary layer connecting the interior with the exterior.}
The amplitude of a fundamental transverse oscillation is greatest at the half-length of the tube, $L/2$, where the shearing is the greatest.
We consider a plane $\Pi$ orthogonal to the tube axis and crossing it at its half-length.
The intersection of this plane with the tube boundary is a circle.
We also assume that the kink oscillation of the magnetic tube is linearly polarised and introduce the angle $\varphi$ in the plane $\Pi$, measured from the direction of the oscillation velocity in the counter-clockwise direction. Then, the shear velocity at the tube boundary takes its maximum at $\varphi = \pi/2$ and $\varphi = 3\pi/2$\/, i.e. at the two points where it is parallel to the oscillation velocity (see Figure~\ref{fig:tube1}).

In order to study the effect of the shearing motions around this region, we model it as a single interface separating temporally periodic counter-streaming flows. We introduce the Cartesian coordinate system $x$, $y$, $z$ with the $x$\/-axis parallel to the direction of the polarisation of the kink oscillation, and the $z$\/-axis parallel to the tube axis. The interior and exterior of the tube are represented by the regions $y < 0$ and $y > 0$, respectively. The equilibrium quantities in these regions are denoted by the subscripts $i$ and $e$\/, respectively.

We assume that the equilibrium magnetic field is in the $xz$\/-plane. Since we wish to obtain the stability criteria both for straight and twisted tubes, we assume that the equilibrium magnetic field is parallel to the $z$-axis in the region $y > 0$, and makes an angle $\theta$ with respect to the $z$-axis in the region $y < 0$. Here, $\theta$ corresponds to the degree of twist (Figure \ref{fig:tube_interface}a), which should be small since highly twisted magnetic flux tubes are prone to other types of instabilities, such as the kink instability, with which we are not concerned in the present study \citep[e.g.][]{Shafranov1958, Kruskal1958, Hood1979}. In the case of a non-twisted flux tube, $\theta = 0$.

In the present model, the background flows are similar to the velocity field at the boundary of a cylindrical flux tube undergoing a transverse oscillation. In transverse oscillations of coronal loops, the displacement of the flux tube boundary is almost perpendicular to the background magnetic field in the low-beta plasma approximation \citep[see, e.g.][]{Ruderman2007}, therefore, we consider unperturbed magnetic fields and flow velocities of the form
\begin{align*}
& \mathbf{B_i} = (B_i \sin \theta, 0, B_i \cos \theta),
\\
& \mathbf{B_e} = (0, 0, B_e),
\\
& \mathbf{U_i} = (U \cos(\Omega t) \cos \theta, 0, - U \cos(\Omega t) \sin \theta),
\\
& \mathbf{U_e} = (- U \cos(\Omega t), 0, 0),
\end{align*}
as illustrated in Figure \ref{fig:tube_interface}b. Here, the period of the oscillatory flow, $2\pi / \Omega$, corresponds to the period of oscillation of the flux tube.

{The kink oscillation of a coronal loop creates not only the oscillating velocity, but also the oscillating magnetic field orthogonal to the background field $\bf B$.
However, in our model we carry out a local analysis of the stability of the region near the middle of the loop where the amplitude of oscillating velocity takes maximum.
Since the oscillating magnetic field has a node at the middle of the loop, that is its amplitude is zero there, we do not take this oscillating magnetic field into account in our model.}

It is worth noting that the problem of oscillatory counter-streaming flows has been previously studied by, e.g. \cite{Kelly1965} and \cite{Roberts1973}.
Our model is an improvement since we do not only consider parallel flows. Furthermore, our model differs from that of \cite{Roberts1973} since we consider magnetic fields perpendicular to the flows on each side of the interface.

We study the dynamics of the outlined problem in the framework of linear ideal MHD.
In the thin flux tube approximation, typically valid for transverse loop oscillations, the effects of compressibility are not significant. As such, we may use the approximation of incompressible plasma, which greatly simplifies the analysis. Thus, the set of governing equations is
\begin{align}
\begin{split}
\label{eq1}
\frac{\mathrm{D} \mathbf{v}}{\mathrm{D} t}
& = - \frac{1}{\rho_{i, e}} \nabla p_T
+ \frac{1}{\mu_0 \rho_{i, e}}( \mathbf{B_{i, e}} \cdot \nabla )\mathbf{b},
\\
\frac{\mathrm{D} \mathbf{b}}{\mathrm{D} t}
& = ( \mathbf{B_{i, e}} \cdot \nabla ) \mathbf{v},
\\
\nabla \cdot \mathbf{v} & = 0,
\\
\nabla \cdot \mathbf{b} & = 0,
\end{split}
\end{align}
where $\mathbf v, \mathbf b$ and $p_T$ are the perturbations of the velocity, magnetic field, and total pressure (magnetic plus plasma), $\rho_{i,e}$ are the background internal and external densities, and $\mu_0$ is the magnetic permeability of free space.
$\mathrm{D}/\mathrm{D} t$ is the material derivative defined by
\begin{equation*}
\dfrac{\mathrm{D}}{\mathrm{D} t}
= \begin{dcases}
\!\begin{aligned}
\dfrac{\partial}{\partial t}
+ U \cos(\Omega t) \cos \theta \dfrac{\partial}{\partial x}
\hphantom{xxxxxxx}
\\
-\: U \cos(\Omega t) \sin \theta \dfrac{\partial}{\partial z}, \quad y < 0,
\end{aligned}
\\
\dfrac{\partial}{\partial t}
- U \cos(\Omega t) \dfrac{\partial}{\partial x}, \qquad \quad y > 0.
\end{dcases}
\end{equation*}
Equation~\eqref{eq1} must be supplemented with the conditions that $p_T$ and $\xi_y$ are continuous at $y = 0$.

We now introduce the Lagrangian displacement $\bm \xi = \bm \xi(\mathbf{x}, t)$, which is related to the velocity perturbation by $\mathbf v (\mathbf{x}, t) = \mathrm{D} \bm \xi / \mathrm{D} t$\/. Combining the momentum and induction equations and substituting the expression for $\mathbf v$ in terms of the displacement yields
\begin{equation}
\label{eq2}
\frac{\mathrm{D}^2 \bm \xi}{\mathrm{D} t^2}
- \frac{1}{\mu_0\rho_{i, e}} ( \mathbf{B_{i, e}} \cdot \nabla )^2 \bm \xi
= - \frac{1}{\rho_{i, e}} \nabla p_T.
\end{equation}
Taking the divergence of this equation, and using $\nabla\cdot\bm \xi = 0$, we obtain Laplace's equation for the total pressure
\begin{equation}
\label{eq3}
\nabla^2 p_T = 0.
\end{equation}
We Fourier-decompose all variables and write them in the form $f = \hat f\exp[i (k_x x + k_z z)]$. We immediately obtain that the solution to Equation~\eqref{eq3} satisfying the condition that it is continuous at $y = 0$ is
\begin{equation}
\label{eq4}
\hat p_T(y) = p_0\left\{\begin{array}{cc} \mathrm{e}^{ky}, & y < 0, \\
\mathrm{e}^{-ky}, & y > 0 , \end{array}\right.
\end{equation}
where $p_0$ is an arbitrary constant, $\mathbf{k} = (k_x, 0, k_z)$ is the wave vector, and $k = \sqrt{k_x^2 + k_z^2}$.

The Fourier-decomposed $y$\/-component of Equation~\eqref{eq2} reads
\begin{align}
\label{eq5}
& \left( \frac{\partial}{\partial t}
+ i k_x U \cos(\Omega t) \cos\theta
- i k_z U \cos(\Omega t) \sin\theta \right)^2 \hat \xi_y
\nonumber \\
& \hphantom{xx} +\: v_{A i}^2 \left( k_x \sin\theta
+ k_z \cos\theta \right)^2 \hat \xi_y
= - \frac{1}{\rho_i} \frac{\partial p_T}{\partial y},
\end{align}
for $y < 0$, and
\begin{equation}
\label{eq6}
\left( \frac{\partial}{\partial t}
- i k_x U \cos(\Omega t) \right)^2 \hat \xi_y
+ v_{A e}^2 k_z^2 \hat \xi_y
= - \frac{1}{\rho_e} \frac{\partial p_T}{\partial y},
\end{equation}
for $y > 0$.
Here, $v_{Ai, e}^2 = B_{i, e}^2 / \mu_0\rho_{i, e}$ are the Alfv\'en speeds on either side of the interface. We substitute Equation~\eqref{eq4} into Equations~\eqref{eq5} and \eqref{eq6}, take $y = 0$, and eliminate the constant $p_0$ from the obtained equations. As a result, we arrive at the equation for the displacement of the boundary,
\begin{align}
\begin{split}
\label{eq7}
& \bigg\{ \frac{\mathrm{d}^2}{\mathrm{d} t^2}
+ 2 i A \cos(\Omega t) \frac{\mathrm{d}}{\mathrm{d} t}
- i \Omega A \sin(\Omega t)
\\
& \hphantom{xx} - B \cos^2(\Omega t) 
+ C \bigg\} \hat \xi_y
= 0, \\
& A
= \frac{U \big[ \rho_i ( k_x \cos\theta - k_z \sin\theta ) 
- \rho_e k_x \big]}{\rho_i + \rho_e},
\\
& B
= \frac{U^2 \big[ \rho_i \left( k_x \cos\theta
- k_z \sin\theta \right)^2
+ \rho_e k_x^2 \big]}{\rho_i + \rho_e},
\\
& C
= \frac{\rho_i v_{A i}^2 \left( k_x \sin\theta
+ k_z \cos\theta \right)^2
+ \rho_e v_{A e}^2 k_z^2}{\rho_i + \rho_e},
\end{split}
\end{align}
where $\hat \xi_y$ is calculated at $y = 0$.

It is now convenient to introduce the magnitude of the wave vector $k$, and the angle between the wave vector and the $x$\/-axis, $\phi$.
We may, then, write
\begin{equation}
\label{eq8}
k_x = k\cos\phi, \quad k_z = k\sin\phi .
\end{equation}
Now, making the variable substitution $\hat \xi_y (t) = g(t) \eta(t)$, where
\begin{equation}
\label{eq9}
g(t) = \exp\left\{- \frac{i A}{\Omega} \sin(\Omega t)\right\},
\end{equation}
we reduce Equation~\eqref{eq7} to
\begin{equation}
\label{eq10}
\frac{\mathrm{d}^2 \eta}{\mathrm{d} \tau^2} + [a - 2 q \cos(2 \tau)] \eta = 0, 
\end{equation}
where
\begin{align}
\begin{split}
\label{eq11}
q & = \frac{r \kappa^2 M_A^2[\cos(\theta + \phi) + \cos\phi]^2}{4 (1 + r)^2},
\\
\alpha & = \frac{\kappa^2[\sin^2(\theta + \phi) + r \bar v_A^2 \sin^2\phi]}{1 + r}, 
\\
a & = \alpha - 2 q,
\end{split}
\end{align}
$\tau = \Omega t$\/, $r = \rho_e / \rho_i$ is the density ratio, $M_A = U / v_{Ai}$ is the Alfv\'en Mach number,  $\bar v_A = v_{Ae} / v_{Ai}$ is the ratio of Alfv\'en speeds, and $\kappa = k v_{Ai} / \Omega$ is the dimensionless wavenumber.

It is important to note that, since $|g(t)| = 1$, the variable substitution does not affect the stability analysis. Hence, unstable perturbations of the boundary correspond to unstable solutions of Equation \eqref{eq10}. Equation~\eqref{eq10} is known as Mathieu's equation \citep{McLachlan1946}. It is interesting that Mathieu's equation also arises in quite a different kind of MHD problem. Namely, it describes the amplification of MHD waves by periodic external forcing \citep[e.g.][]{Zaqarashvili2000,Zaqarashvili2002,Zaqarashvili2005}, and the Rayleigh-Taylor instability of a magnetic interface in the presence of oscillating gravity \citep{Ruderman2018}.

%=============================================================
\section{Investigation of Stability}
\label{sec:stab}
%=============================================================

In this section, we use Equation \eqref{eq10} to study the stability of the tangential discontinuity with an oscillating shear velocity. For comparison, we first briefly outline the well-known results related to the stability of a tangential discontinuity separating steady flows. To the best of our knowledge, these results were first obtained by \cite{Syrovatskii1957} \citep[see also][]{Chandrasekhar1961}.
\vspace*{3mm}

%------------------------------------------------------------------------------
\subsection{Stability of Steady Flows}
\label{subsec:steady}
%------------------------------------------------------------------------------

Before analysing the fully time dependent governing Equation~\eqref{eq10}, we return to Equation \eqref{eq7} and set $\Omega = 0$, in order to perform the analysis of the configuration in the presence of steady flows. Since the coefficients in Equation~\eqref{eq7} are now independent of $t$, we can look for the solution to this equation proportional to $\mathrm{e}^{-i \omega t}$ and obtain the dispersion equation 
\begin{align}
\begin{split}
\label{eq12}
(\rho_i\: + &\: \rho_e) \omega^2
- 2 U k [ \rho_i \cos(\theta + \phi) - \rho_e \cos\phi ] \omega
\\
+ & \: U^2 k^2 [ \rho_i \cos^2(\theta + \phi) + \rho_e \cos^2\phi ]
\\
- & \: \rho_i v_{A i}^2 k^2 \sin^2(\theta + \phi)
- \rho_e v_{A e}^2 k^2 \sin^2\phi
= 0,
\end{split}
\end{align}
where $\omega$ is the angular frequency of the perturbation.

We note that if the roots to Equation~\eqref{eq12} are real, then $\hat \xi_y (t)$ is oscillatory and the system is neutrally stable. However, if complex conjugate roots exist, one of the roots has a positive imaginary part, meaning that $|\mathrm{e}^{-i \omega t}| \to \infty$ as $t \to \infty$\/, and the equilibrium configuration is unstable.
Equation~\eqref{eq12} has complex roots when its discriminant is negative, which occurs when $M_A > M_{A0}$, where 
\begin{equation}
\label{eq13}
M_{A0}^2 = \frac{(1 + r) [ \sin^2(\theta + \phi) + r \bar v_{A}^2 \sin^2\phi ] }
{r [ \cos(\theta + \phi) + \cos\phi ]^2}.
\end{equation}

{The right-hand side of Equation~\eqref{eq13} is singular for $\theta = (2n+1)\pi$ and $\theta + 2 \phi = (2n+1)\pi$\/, where $n$ is any integer number.
The interface is stable for any value of $U$, for $\theta$ and $\phi$ satisfying either of the singularity conditions.
We can see that for $\theta = (2n+1)\pi$, the velocity has the same magnitude and direction on both sides of the interface, meaning that there is no velocity jump across the interface.
Hence, the equilibrium is static in the reference frame moving with the speed $U$ in the positive $x$-direction and, consequently, the presence of flow does not cause instability.
In the second case, the interface is stable with respect to perturbations having wave vectors defined by $\phi = -\frac12\theta + \left(n+\frac12\right)\pi$.
The projection of the velocity on these wave vectors is the same on both sides of the interface, that is, there is no jump in the velocity projection across the interface.
Hence, these perturbations are stable for any value of $U$.} 

The Alfv\'en Mach number, $M_{A0}$, takes its minimum value with respect to $\phi$ at $\phi = \phi_0$\/, where
\begin{equation}
\label{eq14}
\phi_0 = -\arctan \left(\frac{ \sin\theta}{\cos\theta + r \bar v_{A}^2}\right).
\end{equation}
Substituting Equation~\eqref{eq14} into Equation \eqref{eq13}, we obtain
\begin{equation}
\label{eq15}
\min \{ M_{A0}^2 \} = \frac{\bar v_{A}^2 (1 + r)\tan^2(\theta/2)}{1 + r \bar v_{A}^2} .
\end{equation}
It follows that the system is stable for any value of $M_A$ below $\min\{M_{A0}\}$, while there are always unstable perturbations when $M_A > \min\{M_{A0}\}$.
{Equation~\eqref{eq15} suggests there are no stable perturbations for $\theta = 0$, and is in angreement with \cite{Syrovatskii1957}.}

We note that the instability growth rate is proportional to $k$, which implies that the growth rate tends to infinity as $k \to \infty$.
Since the growth rate is unbounded, we say that the initial value problem describing the evolution of the surface of discontinuity is ill-posed.
This behaviour is further studied in Section \ref{subsec:ivp}.
\vspace*{3mm}

%------------------------------------------------------------------------------
\subsection{Stability of Oscillating Flows}
\label{subsec:oscillating}
%------------------------------------------------------------------------------

We now use Equation~\eqref{eq10} to study the stability for arbitrary values of the equilibrium quantities. Floquet's theorem states that Equation~\eqref{eq10} has a solution of the form
\[
\eta_+(\tau) = \mathrm{e}^{\mu \tau} P(a, q,\tau) ,
\]
where $\mu = \mu(a, q)$ is the characteristic exponent, and $P(a, q,\tau)$ is a periodic function in $\tau$, with period $\pi$ \citep[see, e.g.,][]{McLachlan1946, Abramowitz1965}. Since Equation~\eqref{eq10} is invariant with respect to the substitution $-\tau \to \tau$ it follows that $\eta_-(\tau) = \mathrm{e}^{-\mu \tau} P(a, q,-\tau)$ is also a solution to this equation. Then, the general solution to Equation~\eqref{eq10} is the linear combination of $\eta_+(\tau)$ and $\eta_-(\tau)$ unless $i\mu$ is an integer number.

The parameter $\mu$ determines the nature of solutions to Mathieu's equation. 
We may always assume that $\Re(\mu) > 0$, unless $\mu$ is purely imaginary, where $\Re$ indicates the real part of a quantity. Since we may write 
\[
\mathrm e^{\mu \tau} = \exp(\Re(\mu) \Omega t)\exp(i\Im(\mu) \Omega t),
\]
where $\Im$ indicates the imaginary part of a quantity, it follows that purely imaginary values of $\mu$ correspond to neutrally stable solutions, while real and complex values correspond to unstable solutions. Hence, $\Re(\mu) > 0$ corresponds to an unstable perturbation. Unfortunately, $\mu$ cannot be easily computed analytically, and, for this reason, we perform a numerical analysis to gain further insight.

\begin{figure*}[t]
\gridline{\fig{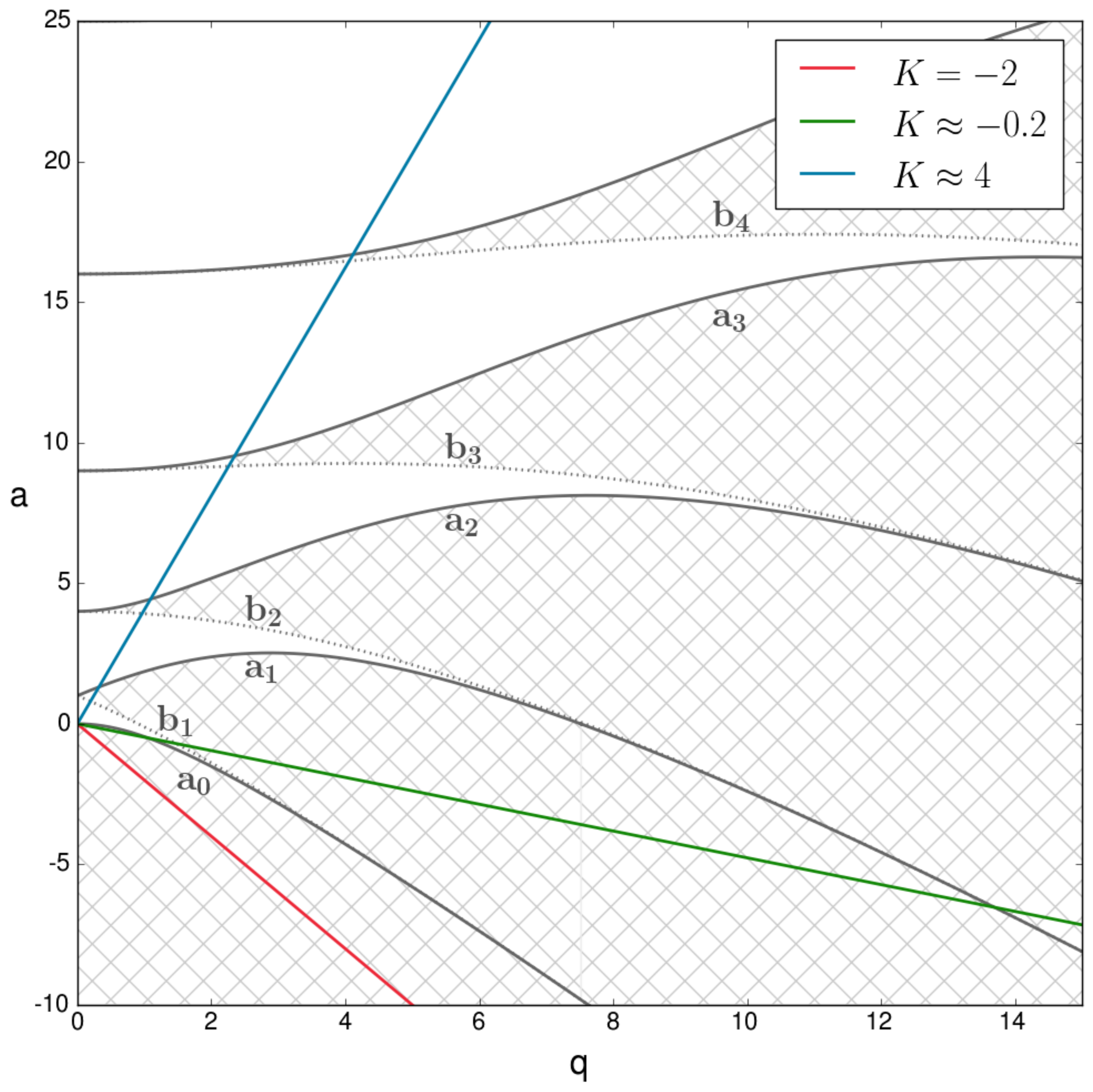}{0.85\columnwidth}{(a)}
\hfill
	     \fig{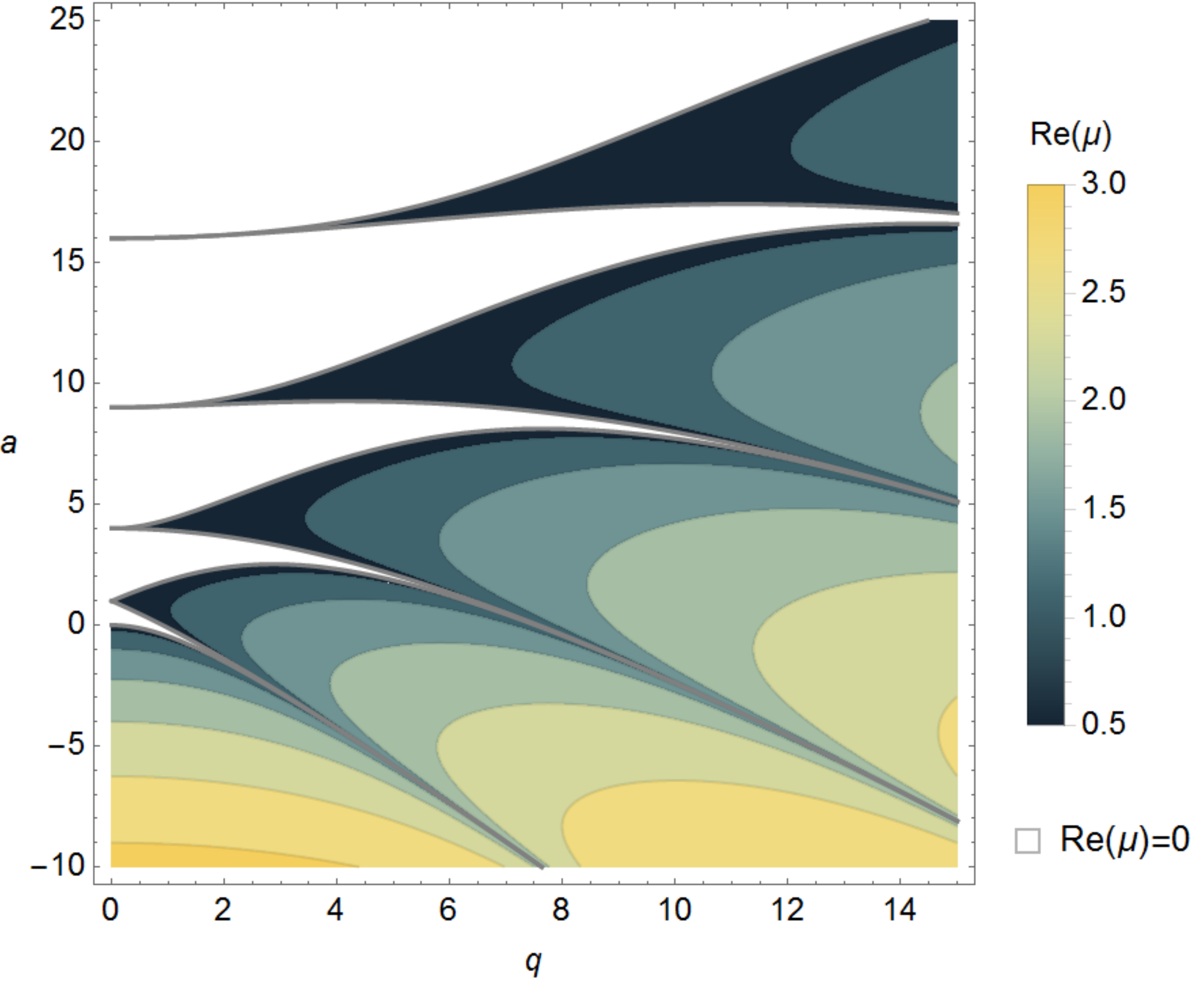}{\columnwidth}{(b)}}
\caption{
The stability diagram for solutions to Mathieu's equation (left panel). Solutions are stable/unstable for $(q, a)$ in the white/hatched region. The curves $a = a_j(q)$ and $a = b_j(q)$ are shown by solid and dotted lines, respectively. The blue, green, and red straight lines correspond to $K \approx 4$, $K \approx - 0.2$, and $K = -2$, respectively. In the panel on the right, the real part of $\mu$ is plotted for $q > 0$.
\label{fig:stability_diagram}
}
\end{figure*}

Following \cite{McLachlan1946}, we plot the stability diagram of Equation~\eqref{eq10} in the $qa$\/-plane (Figure~\ref{fig:stability_diagram}a). In accordance with the definition of $q$ in Equation \eqref{eq11}, we only consider $q > 0$. The white and hatched regions correspond to purely imaginary and real/complex values of $\mu$, respectively, and thus, to stable and unstable solutions to Equation~\eqref{eq10}.
The contours bounding the regions are defined by the condition that $i\mu$ is an integer number, so that Equation~\eqref{eq10} has either $\pi$ or $2\pi$\/-periodic solutions when the point $(q,a)$ is on one of these contours. These contours are called the characteristic curves, and are defined by the equations $a = a_j(q)$ and $a = b_j(q)$. These functions satisfy the inequalities $a_j < b_{j+1} < a_{j+1}$\/, where $j = 0,1,2,\dots$ The curves $a_j(q)$ and $b_j(q)$ are shown by solid and dotted lines, respectively, in Figure~\ref{fig:stability_diagram}a. The asymptotic behaviour of $a_j(q)$ and $b_{j+1}(q)$ for large $q$ is given by $a_j(q) \sim b_{j+1}(q) \sim -2q$ \citep{Abramowitz1965}.

Complementary to the above, Figure~\ref{fig:stability_diagram}b shows the values of the characteristic exponent $\mu$. Purely imaginary solutions are plotted in white, and are separated from real/complex solutions by the characteristic curves, while the real part of $\mu$ is plotted in contours in the unstable regions.

The coefficients in Equation~\eqref{eq10} depend on six dimensionless parameters.
Four of these parameters, $r$\/, $\theta$\/, $M_A$\/, and $\bar v_A$\/, are only dependent on the equilibrium quantities, while the other two, $\kappa$\/, and $\phi$\/, are related to particular perturbations, and are thus arbitrary. Hence, we must study the behaviour of solutions to Equation~\eqref{eq10} for all possible values of these two parameters. It is also straightforward to see that $q$ and $a$ are invariant with respect to the substitution $\phi + \pi \to \phi$\/. This enables us to only consider values of $\phi$ in the interval $[-\pi/2, \pi/2]$.

We now wish to study the behaviour of solutions to Equation \eqref{eq10} for arbitrary $\kappa$. We begin by noting that, when $\phi$ is fixed and $\kappa$ varies from 0 to $\infty$ we obtain a straight line in the $qa$\/-plane. Using Equations~\eqref{eq11}, the equation of this line may be written as 
\begin{equation}
\label{eq16}
a = Kq, \quad K = \frac{4 M_{A 0}^2}{M_A^2} - 2.
\end{equation}
From Equations \eqref{eq13} and \eqref{eq16}, we note that $K > -2$ for any $\theta \neq 0$ and any values of the other parameters. Considering the asymptotic behaviours of the characteristic curves, it follows that the line $a = K q$ always intersects all curves $a = a_j(q)$ and $a = b_{j+1}(q)$, for $j = 0,1,\dots$ Hence, there always exist some values of $\kappa$ and $\phi$ for which perturbations are unstable, regardless of the values of the other parameters. This implies that the tangential discontinuity separating oscillating flows is unstable for any value of $M_A$, which is qualitatively different from the discontinuity separating steady flows considered in Subsection \ref{subsec:steady}.
In the case of no magnetic shearing, when $\theta = 0$, perturbations with $\phi = 0$ and any $\kappa$ are unstable since the line $a = K q$ will always be under the curve $a_0(q)$. This is illustrated by the red line in Figure \ref{fig:stability_diagram}a.
The straight lines in Figure \ref{fig:stability_diagram}a are further discussed in Subsection \ref{subsec:sigma}.

%------------------------------------------------------------------------------
\subsection{The Initial Value Problem}
\label{subsec:ivp}
%------------------------------------------------------------------------------

We now consider the initial value problem for Equation~(\ref{eq10}).
We fix $\phi$ and study how the properties of the initial value problem depend on $M_A$\/. First, we consider $M_A > M_{A0}(\phi)$, which, implies that $K < 2$ due to Equation \eqref{eq16}, and we prove that, in this case, the instability growth rate is unbounded. Let us introduce the scaled variables $\tilde{a} = \kappa^{-2} a$\/, $\tilde{q} = \kappa^{-2} q$\/, and $\tilde{\tau} = \kappa\tau$, and rewrite Equation~\eqref{eq10} as
\begin{equation}
\label{eq17}
\frac{\mathrm{d}^2 \eta}{\mathrm{d}\tilde{\tau}^2} + [\tilde{a} - 
2\tilde{q}\cos(2\tilde{\tau}/\kappa)] \eta = 0. 
\end{equation}
It is important to note that $\tilde{a}$ and $\tilde{q}$ are independent of $\kappa$\/, and $\tilde{a} = K\tilde{q}$\/. We consider this equation on the interval $\tilde{\tau} \in [0,\tilde{\tau}_0]$, where $\tilde{\tau}_0 = \kappa\arcsin h$ and $h = \frac12\sqrt{1 - K/2}$. Since $K>-2$, it follows that
\begin{equation}
\label{eq18}
2\tilde{q}\cos(2\tilde{\tau}/\kappa) - \tilde{a} \geq 4h^2\tilde{q},
\end{equation}
for $\tilde{\tau} \in [0,\tilde{\tau}_0]$.

We now consider equation
\begin{equation}
\label{eq19}
\frac{\mathrm{d}^2 \eta}{\mathrm{d}\tilde{\tau}^2} - 4h^2\tilde{q}\eta = 0, 
\end{equation}
and a solution to this equation 
\begin{equation}
\label{eq20}
\eta_1 = \eta_0\exp(2h\tilde{q}^{1/2}\tilde\tau) = 
\eta_0\exp\big(\tau\sqrt{q(1 - K/2)}\big),
\end{equation}
where $\eta_0$ is an arbitrary constant. This solution satisfies the initial conditions
\begin{equation}
\label{eq21}
\eta_1 = \eta_0, \quad \frac{\mathrm{d}\eta_1}{\mathrm{d}\tilde{\tau}} =
2h\eta_0\tilde{q}^{1/2} \quad \mbox{at} \;\; \tilde{\tau} = 0.
\end{equation}
We also consider a solution $\eta_2$ to Equation~(\ref{eq17}) satisfying the same initial conditions. Then, it follows from Equation~(\ref{eq18}) and the comparison theorem \citep[e.g.][]{Coddington1955} that $\eta_2 \geq \eta_1$ for $\tilde{\tau} \in [0,\tilde{\tau}_0]$. The initial condition for $\eta_2$ can be rewritten as
\begin{equation}
\label{eq22}
\eta_2 = \eta_0, \quad \frac{\mathrm{d}\eta_2}{\mathrm{d}\tau} =
2h\eta_0\kappa^{-1}\tilde{q}^{1/2} \quad \mbox{at} \;\; \tilde{\tau} = 0.
\end{equation}
This result implies that $\eta_2$ and $\mathrm{d}\eta_2/\mathrm{d}\tau$ are bounded at $\tau = 0$ for $\kappa \in (0,\infty)$. Then, it follows from the inequality $\eta_2 \geq \eta_1$ and Equation~(\ref{eq20}) that, for any $\tau_0 \in (0,\arcsin h)$, there is such a solution to Equation~(\ref{eq17}) that it is bounded together with its first derivative at $\tau = 0$ for any value of $\kappa$\/, but it is unbounded at $\tau = \tau_0$ as $\kappa \to \infty$\/. Hence, the instability growth rate is unbounded. This result implies that the initial value problem describing the evolution of the perturbed discontinuity is ill-posed when $M_A > \min \{ M_{A0} \}$.

Now, we assume that $M_A < M_{A0}(\phi)$, so that, in accordance with Equation \eqref{eq16}, $K > 2$ and $a > 2q$. We calculate the instability increment for $\kappa \gg 1$. Let $\bar\eta(\tau)$ be the solution to Equation~\eqref{eq10}, satisfying the initial conditions
\begin{equation}
\label{eq23}
\bar\eta = 1, \quad \frac{\mathrm{d}\bar\eta}{\mathrm{d}\tau} = 0 \quad
\mbox{at} \quad \tau = 0. 
\end{equation}
Then, the characteristic exponent is defined by the equation \citep{Abramowitz1965}
\begin{equation}
\label{eq24}
\cosh(\pi\mu) = \bar\eta(\pi).
\end{equation}
We use the WKB method and look for a solution to Equation~\eqref{eq10} in the form  $\eta_+ = \mathrm{e}^{\kappa\Theta}$\/. Substituting this expression into Equation~\eqref{eq10} we obtain
\begin{equation}
\label{eq25}
\kappa^{-1}\frac{\mathrm{d}^2\Theta}{\mathrm{d}\tau^2} + 
\left(\frac{\mathrm{d}\Theta}{\mathrm{d}\tau}\right)^2 +
\tilde{a} - 2\tilde{q}\cos(2\tau) = 0. 
\end{equation}
We impose the condition $\Theta = 0$ at $\tau = 0$. Then, we look for the solution to this equation in the form of expansion
\begin{equation}
\label{eq26}
\Theta = \Theta_1 + \kappa^{-1}\Theta_2 + \dots 
\end{equation}
Substituting this expansion into Equation~\eqref{eq25} and collecting terms of the order of unity we obtain
\begin{equation}
\label{eq27}
\left(\frac{\mathrm{d}\Theta_1}{\mathrm{d}\tau}\right)^2 =
2\tilde{q}\cos(2\tau) - \tilde{a}. 
\end{equation}
The solution to this equation satisfying the condition $\Theta_1 = 0$ at $\tau = 0$ is
\begin{equation}
\label{eq28}
\Theta_1 = i\int_0^\tau\sqrt{\tilde{a} - 2\tilde{q}\cos(2\tau')}\,\mathrm{d}\tau',
\end{equation}
where we chose the plus sign at the square root. 

In the next order approximation we collect terms of the order of $\kappa^{-1}$ in Equation~\eqref{eq25} to obtain
\begin{equation}
\label{eq29}
\frac{\mathrm{d}^2\Theta_1}{\mathrm{d}\tau^2} + 
\frac{\mathrm{d}\Theta_1}{\mathrm{d}\tau}
\frac{\mathrm{d}\Theta_2}{\mathrm{d}\tau} = 0. 
\end{equation}
Using Equation~\eqref{eq28} we find that the solution to this equation satisfying the condition $\Theta_2 = 0$ at $\tau = 0$ is
\begin{equation}
\label{eq30}
\Theta_2 = -\frac12\ln\frac{\tilde{a} - 2\tilde{q}\cos(2\tau)}{\tilde{a} - 2\tilde{q}}.
\end{equation}
Recall that $\eta_-(\tau) = \eta_+(-\tau)$ is also a solution to Equation~\eqref{eq10}. Then, since $\Theta_1(\tau)$ is an odd function and $\Theta_2(\tau)$ is an even function, it follows that 
\begin{equation}
\label{eq31}
\bar\eta = \frac{\eta_+ + \eta_-}2 = 
\mathrm{e}^{\Theta_2}\cos(\kappa\Theta_1) + {\cal O}\big(\kappa^{-1}\big).
\end{equation}
Introducing the notation $\chi = \Theta_1(\pi)$ and $\gamma = \Theta_2(\pi)$ we transform Equation \eqref{eq10} to
\begin{equation}
\label{eq32}
\cosh(\pi\mu) = \mathrm{e}^\gamma\cos(\kappa\chi).
\end{equation}
When the absolute value of the right-hand side of this equation does not exceed unity the two values of $\mu$ satisfying this equation are purely imaginary and the corresponding wave mode is neutrally stable. When the absolute value of the right-hand side is larger than unity one of the two values of $\mu$ satisfying this equation has positive real part and the corresponding wave mode grows exponentially. However, we can observe that the right-hand side of Equation~\eqref{eq32} is bounded for any $\kappa$\/. This implies that the real part of $\mu$ is also bounded, and the same is true for the growth rate. We made this conclusion for a particular value of $\phi$ and $M_A < M_{A0}(\phi)$. If we now assume that $M_A < \min \{ M_{A0} \}$, then the growth rate of any wave mode is bounded. This means that the initial value problem describing the evolution of the discontinuity is well-posed when $M_A < \min \{ M_{A0} \}$. From Equation \eqref{eq15} we see that this condition may be written in the approximate form as
\begin{equation}
\label{eq33}
M_A < \frac{\bar v_A\theta}2\sqrt{\frac{1 + r}{1 + r \bar v_A^2}}, 
\end{equation}
since, typically, $\theta \ll 1$.

%============================================================
\section{Application to Transverse Coronal Loop Oscillations}
\label{sec:loop}
%============================================================

The aim of this section is twofold. First, we further elaborate the analysis of Section~\ref{sec:stab} by considering the $\sigma$-stability of Equation \eqref{eq10}.
Afterwards, we apply some of the results obtained in Subsections~\ref{subsec:oscillating} and \ref{subsec:ivp} to the stability of coronal loop oscillations.

%------------------------------------------------------------------------------
\subsection{The $\sigma$-stability}
\label{subsec:sigma}
%------------------------------------------------------------------------------

We now use the concept of $\sigma$-stability, first introduced by \cite{Goedbloed1974} and \cite{Sakanaka1974}. This concept is used in studies of thermonuclear plasma confinement where it is necessary that perturbation amplitudes remain sufficiently small on some relevant time scale. An equilibrium is $\sigma$-stable if the amplitudes of unstable perturbations grow at most like $\exp(\sigma t)$.

We apply the concept of $\sigma$-stability to the analysis of the KH instability induced by transverse oscillations of solar coronal loops.
We say that a transverse coronal loop oscillation is $\sigma$-stable if the growth time of the KH instability exceeds the damping time due to resonant absorption.
{It is important to note that, in this paper, we only consider the KH instability due to the transverse oscillation of coronal loops without a transitional layer.
If a transitional layer is present, the KH instability may still occur in coronal loops after the transverse oscillation is damped \citep{Terradas2018} as a result of increased shearing motions due to resonant absorption \citep{Heyvaerts1983, Browning1984}.}

Let $t_D = \alpha P$ be the damping time, where $P = 2 \pi / \Omega$ is the oscillation period, and $\alpha$ varies from 1 to 5 \citep[see, e.g.,][]{Goddard2016}. It follows from our definition that $\sigma = 1/ \Omega t_D$, or
\begin{equation}
\label{eq34}
\sigma = \frac{1}{2\pi\alpha}.
\end{equation}
When $\alpha$ varies from 1 to 5, $\sigma$ decreases from approximately 0.16 to 0.03.
We see that, in any case, the interface cannot be $\sigma$\/-stable if the maximum growth rate exceeds 0.16, which implies that if the interface is $\sigma$\/-stable then the increment is much less than unity. It is shown in Appendix~\ref{sec:appendix} that, in this case, the maximum growth rate for fixed $\phi$ is approximately equal to $1/2K$. Then, the maximum growth rate for all values of $\phi$ is $1/2K_m$\/, where $K_m = \min_\phi K$\/. Hence, the $\sigma$\/-stability condition reads
\begin{equation}
\label{eq35}
K_m \geq \frac1{2\sigma}, \quad K_m = \frac{4\min\{M_{A 0}^2\}}{M_A^2} - 2.
\end{equation}
To estimate $K_m$ we take as typical values $r = 1/3$ and $\bar v_A^2 = 3$. Then, using Equations~\eqref{eq15} and \eqref{eq35}, and taking into account the fact that, typically, $\theta \ll 1$, we reduce the $\sigma$-stability criterion to
\begin{equation}
\label{eq36}
\theta \geq \frac{M_A}{2} \sqrt{4 + \frac1\sigma}.
\end{equation}

The typical displacement of a kink-oscillating coronal loop is of the order of the loop radius.
Then, the ratio of the velocity to $v_{Ai}$ is of the order of the loop radius and length. Hence, the typical value is $M_A = 0.01$.
It follows from Equation~\eqref{eq36} that the interface is $\sigma$\/-stable if $\theta \gtrsim 1^\circ$ for $\alpha = 1$, and $\sigma$\/-stable if $\theta \gtrsim 2^\circ$ for $\alpha = 5$.
{Similar to \cite{Terradas2018} we define the number of turns of a magnetic field as 
$$
N_{tw} = \frac{L B_\phi}{2\pi RB_z},
$$
where $B_\phi$ and $B_z$ are the azimuthal and axial components of the magnetic field in cylindrical coordinates with the $z$\/-axis coinciding with the loop axis, and $R$ is the radius of the loop cross-section. Now we use the relation $B_\phi/B_z = \theta$ valid for small $\theta$ and $R/L = 100$ as a typical value for coronal loops.}
We obtain that even the maximum value $\theta = 2^\circ$ corresponds to only about a half-turn of magnetic field lines from one loop footpoint to the other.
Hence, the loop boundary is $\sigma$-stable even for a very moderate magnetic twist.
\begin{figure}[t]
\centering
\includegraphics[width=0.99\columnwidth]{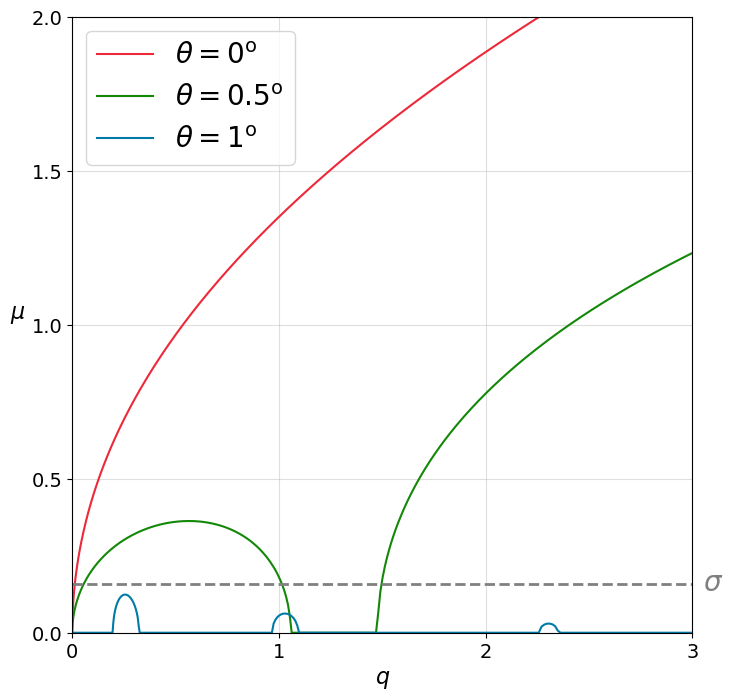}
 \caption{The growth rate of the instability, $\mu$, plotted with respect to $q$. The red, green, and blue lines correspond to the lines in Figure \ref{fig:stability_diagram}a}
 \label{fig:mu}
\end{figure}

In Figure \ref{fig:mu}, we present the values of $\mu$ associated with the three straight lines in Figure \ref{fig:stability_diagram}. We assumed that $r = 1/3$, $\bar v_A^2 = 3$, $M_A = 0.01$, and $\phi = \phi_0$ so that $K = K_m$\/. For $\theta = 0$, $\mu$ is a monotonically increasing function of $\kappa$, and perturbations with any $q$ are unstable. The green curve corresponds to $\theta = 0.5^\circ$, and is unbounded as $\kappa \to \infty$ since $\min M_{A0} \approx 0.0062 < M_A$. Finally, the blue curve, which corresponds to $\theta = 1^\circ$, is bounded for $\kappa \in (0,\infty)$ since $\min M_{A0} \approx 0.0123 > M_A$\/. The equation of the dashed line is $\mu = 0.16$, and we see that the loop with $\theta = 1^\circ$ is $\sigma$-stable for $\sigma$ defined in Equation~\eqref{eq34} with $\alpha = 1$.

We note that if a magnetic loop is $\sigma$\/-stable, then the initial value problem describing the evolution of its boundary perturbation is well-posed. However, the converse is not always true. The initial value problem is well-posed if the growth rate is bounded, but it may still be very large. On the other hand, a magnetic loop is $\sigma$\/-stable when the maximum growth rate is below a definite and, usually, sufficiently small number.

\subsection{The $\sigma$-stability in Numerical Models}

{
We compare our results with those of \cite{Howson2017b} and \cite{Terradas2018}, who studied numerical models of the TWIKH instability in twisted magnetic flux tubes.
Both models consider flux tubes with a finite-width transitional layer, where the density decreases from a high value in the core region of the flux tube to a low value in the surrounding plasma.
The presence of the transitional layer results in damping of kink oscillations due to resonant absorption, such that the concept of $\sigma$-stability is applicable.
Since we do not consider the effects of resonant absorption in the present work, we  may only make a qualitative comparison between results.
}

{
\cite{Howson2017b} considered both twisted and untwisted tubes, subject to a transverse oscillation with a period of the fundamental mode of 280\,s.
Both the oscillation period and damping time were practically unaffected by the magnetic twist.
Using the dependence of the oscillation amplitude on time presented in \cite{Howson2017b}, we estimate that the damping time of the transverse oscillation was approximately 1000\,s.
We also estimate that the instability growth time increases from approximately 600\,s in the case of the untwisted tube to approximately 700\,s when the twist is maximal, which signals a relatively weak dependence of damping time on the degree of twist.
The increase in growth time with increase in twist qualitatively agrees with the results obtained in the present work.
}

{
We have shown in the previous subsections that, in a tube with a sharp boundary (i.e. no transitional layer), the instability growth time is zero.
Therefore, it is clear that the presence of a transitional layer strongly reduces the instability growth rate, and, in the model studied by \cite{Howson2017b}, the effect of the transitional layer on the instability increment is stronger than the effect of twist.
Since the damping time was larger than the instability growth time, the oscillations studied by \cite{Howson2017b} were $\sigma$\/-unstable for all values of twist.
}

{
\cite{Terradas2018} also studied kink oscillations of twisted tubes with transitional layer of thickness $l$.
They considered three values of the transitional layer thickness, $l/R = 0.3,\,1$, and 2, where $R$ is the tube radius.
They also considered several values of the magnetic twist, with the turn of magnetic lines varying from 0 (no twist) to 1.65 turns.
}

{
Similarly to \cite{Howson2017b}, \cite{Terradas2018} obtained that the damping time is practically independent of the twist.
It was approximately equal to $4P$ for $l/R = 0.3$, where $P$ is the oscillation period.
They did not give the value of damping time for other values of the transitional layer thickness.
However, since \cite{Terradas2018} obtained that the numerically calculated values of damping time agree very well with those given by the analytical expression, we can use the fact that the damping time is inversely proportional to $l/R$.
We obtain the estimates that the damping time is about $1.2P$ for $l/R = 1$ and $0.6P$ for $l/R = 2$.
Even if we underestimated the damping time, then the first time is definitely less than $2P$, and the second one is less than $P$.
}

{
The authors also estimated the instability growth time.
They obtained that it strongly depends on the degree of twist.
For $l/R = 0.3$ it increases from about $1.5P$ to about $3P$ when the turn of magnetic field lines varies from 0 to 1.65.
Hence, it is always smaller than the damping time meaning that the oscillations are $\sigma$-unstable.
When $l/R = 1$, the instability growth time increases from about $2.5P$ to about $7.5P$\/.
Finally, when $l/R = 2$ the instability growth time is about $5P$ when there is no twist, and quickly becomes larger than $10P$ when the twist increases.
Hence, the oscillations are always $\sigma$\/-stable when $l/R = 1$ and $l/R = 2$. Since they are $\sigma$\/-stable even when there is no twist, it is obvious that there is a substantial contribution of the transitional layer in the reduction of the instability increment.
However, it is also obvious that the twist substantially contributes in this reduction.
}     

%--------------------------------------------------------------
\subsection{Coronal Loop Parameters}
\label{subsec:loop}
%--------------------------------------------------------------

\begin{figure*}[t]
\gridline{\fig{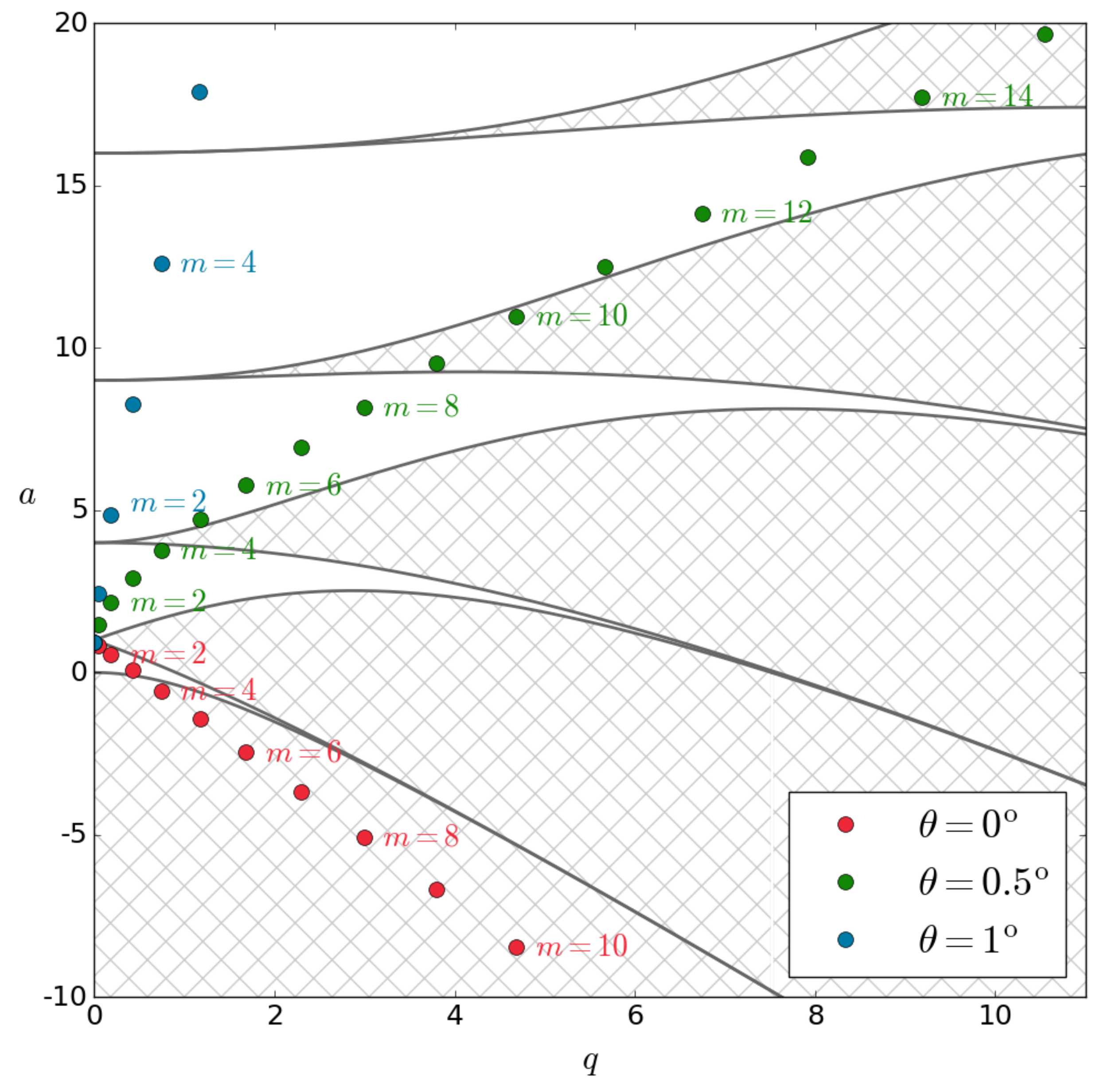}{\columnwidth}{(a)}
\hfill
	     \fig{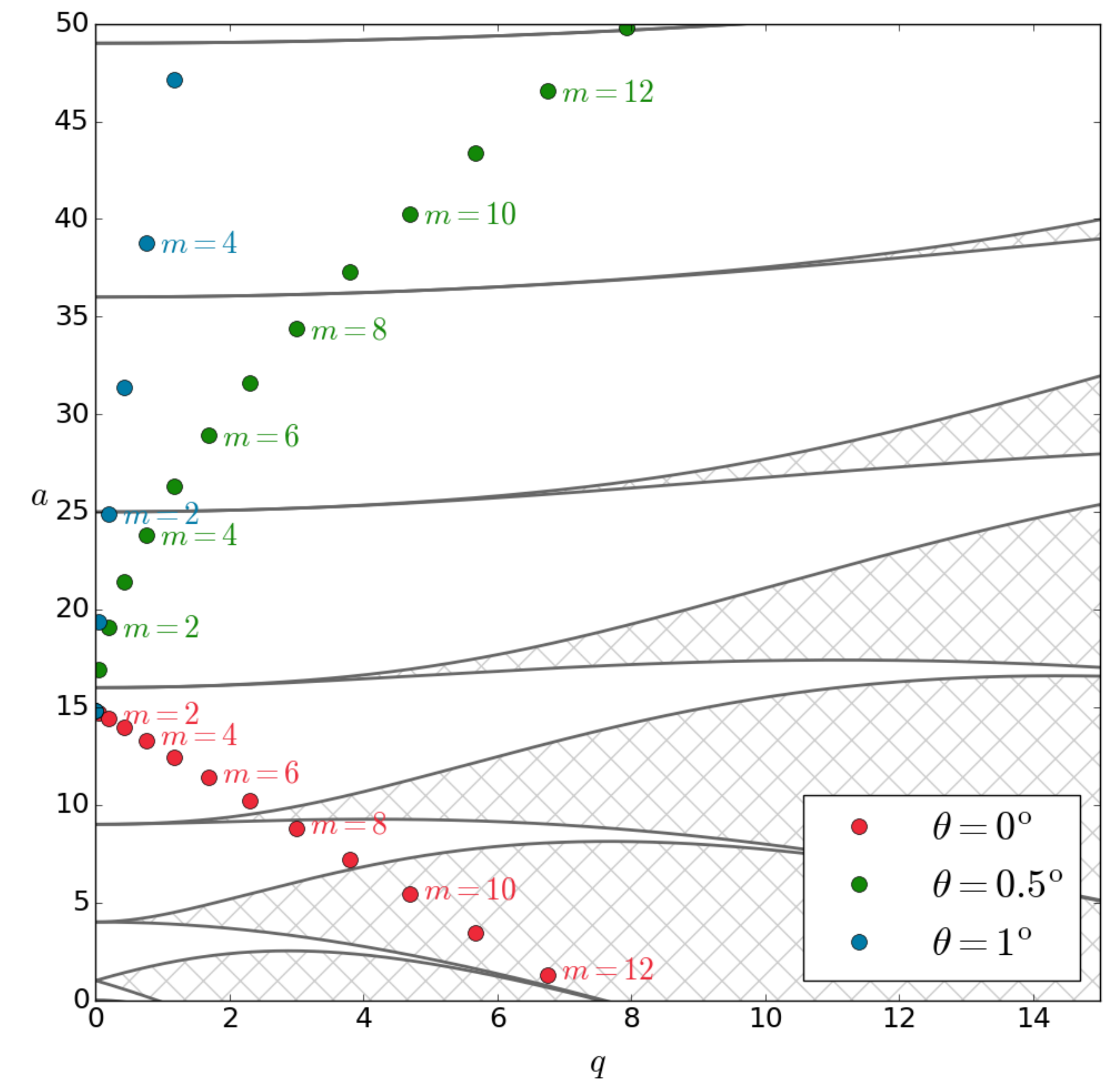}{\columnwidth}{(b)}}
\caption{
The dependence of the growth rate on $m$ for $M_A = 0.01$, $r = 1/3$, $\bar{v}_A^2 = 3$, $n=1$ (left) and $n=4$ (right). The red, green and blue dots correspond to increasing degrees of twist.
\label{fig:stability_loop}
}
\end{figure*}

The model that we outlined in the previous sections can be only applied for the local analysis of the stability of the boundary of an oscillating magnetic tube. In this analysis, we can consider oscillations with the characteristic scale in the azimuthal direction that is much smaller than the tube radius $R$\/, and   the characteristic scale in the axial direction that is much smaller than the tube length $L$\/. Hence, we take 
\begin{equation}
\label{eq37}
k_x = \frac mR, \quad k_z = \frac{\pi n}L,
\end{equation}
where $m$ and $n$ are sufficiently large integer numbers. Using Equations~\eqref{eq8} and \eqref{eq37} we obtain 
\begin{equation}
\label{eq38}
k^2 = \frac{m^2}{R^2} +  \frac{\pi^2 n^2}{L^2}, \quad 
\tan\phi = \frac{\pi nR}{mL} . 
\end{equation}
We assume that $n \lesssim |m|$. Since in coronal magnetic loops $R \ll L$, it follows that we may use the approximate expressions
\begin{equation}
\label{eq39}
k \approx \frac{|m|}R , \quad \phi \approx \frac{\pi nR}{mL} . 
\end{equation}
Throughout this section we assume that $\bar v_A^2 = r^{-1}$. This assumption holds if the magnitudes of the interior and exterior magnetic fields are equal, which is typically true for coronal loops. We also assume that $\theta \ll 1$. Then, we obtain the approximate expressions
\begin{equation}
\label{eq40}
M_{A 0}^2 = \frac{1 + r}{4r}\bigg[\left(\theta + \frac{\pi nR}{mL}\right)^2 +
\frac{\pi^2 n^2 R^2}{m^2 L^2}\bigg] ,
\end{equation}
\begin{equation}
\label{eq41}
\min \{ M_{A0}^2 \} = \frac{(1 + r)\theta^2}{8r} .
\end{equation}
The condition $M_A^2 <  \min \{ M_{A0}^2 \}$ gives 
\begin{equation}
\label{eq42}
\theta > M_A\sqrt{ \frac{8r}{1 + r}} .
\end{equation}
If we take $r = 1/3$, the right-hand side of this inequality is approximately equal to $M_A$, that is it is of the order of 0.01. Hence, the inequality~\eqref{eq42} can be satisfied even for quite moderated twist. If the inequality is satisfied, then the IVP describing the evolution of the tube boundary is well-posed and the growth rate of perturbations is bounded.

In Figures~\ref{fig:stability_loop}, we show the dependence of the growth rate on $m$ for $n = 1$ (left) and $n=4$ (right), $M_A = 0.01$, $r = 1/3$, $\bar{v}_A^2 = 3$, $R/L = 200$, and $\theta = 0^\circ$ (red), $\theta = 0.5^\circ$ (green) and $\theta = 1^\circ$ (blue). We note that, obviously, $n = 1$ does not satisfy the condition that $n$ is large, so we considered $n = 1$ only for comparison. While, for $n=1$, the points in the $qa$\/-plane corresponding to $\theta = 0^\circ$ are virtually unchanged as compared to the line in Figure \ref{fig:stability_diagram}, for $n=4$ they are shifted upwards considerably.
This is also the case for $\theta = 0.5^\circ$. We see that for $n=1$ there are some modes which are unstable in the range selected, for $n=4$ there are no such modes.
There may be unstable modes for $\theta = 0.5^\circ$ and $n=4$, but only for very large $m$\/. In terms of the IVP, for $\theta = 1^\circ$, corresponding to a well-posed solution, no value of $m$ corresponds to an unstable solution in the $qa$\/-plane. In general, well-posed solutions seem to be unstable only for very large $m$\/. These results are significant since they suggest that very localised longitudinal perturbations of the flux tube are generally more stable.

%============================================================
\section{Summary and Discussion}
\label{sec:sum}
%============================================================

In this work, we performed the first {local stability analysis} of the transverse wave induced Kelvin-Helmholtz instability of twisted solar coronal loops.
We modelled the region on the loop boundary where the shear flows are the greatest as a tangential discontinuity separating time-periodic counter-streaming flows.
To model the magnetic twist in coronal loops we assumed that the equilibrium magnetic fields on either side of the discontinuity are not parallel.
The flow velocities at the two sides of the discontinuity have opposite directions and equal magnitudes oscillating harmonically.
For the sake of mathematical simplicity, we assumed that the plasma on both sides of the interface is incompressible.
Using the linearised set of ideal MHD equations, we derived the governing equation describing the evolution of the shape of the tangential discontinuity, known as Mathieu's equation.

We employed Mathieu's equation to study the stability of the discontinuity.
For comparison, we first presented the results of the stability analysis in the case of steady flows, which we obtained by setting the flow oscillation frequency to zero.
In this case, the stability of the discontinuity is determined by the Alfv\'en Mach number, which is defined as the ratio of the background velocity magnitude to the Alfv\'en speed at one side of the interface.
The discontinuity is unstable when the Alfv\'en Mach number exceeds a critical value, and the instability growth rate is proportional to the wavenumber, and thus unbounded.
This implies that the initial value problem describing the evolution of the perturbed discontinuity is ill-posed.
We note that the critical Alfv\'en number is zero when there is no magnetic shear. 

In contrast to the interface separating steady flows, the tilted magnetic field cannot stabilise the discontinuity if the flows oscillate. A similar result was obtained by \cite{Roberts1973} in the case of MHD tangential discontinuity with the magnetic field having the same direction at both sides and the flow velocity parallel to the magnetic field.

Even though the interface is always unstable, the critical Alfv\'en Mach number still plays an important role in the stability properties. We showed that the growth rate of the instability is unbounded when the Alfv\'en Mach number exceeds the instability threshold, and thus the initial value problem is ill-posed. Hence, in this case the stability properties are qualitatively the same as in the case of steady flows. On the other hand, when the the Alfv\'en Mach number is below its critical value, the instability increment is bounded, and the initial value problem is well-posed.

In Section \ref{subsec:sigma}, we introduced the definition of $\sigma$-stability for kink oscillating coronal loops, which states that the loop is $\sigma$-stable if the growth time of the instability exceeds the resonant damping time of the transverse oscillation.
We obtained the criterion for the $\sigma$-stability and showed that, for parameters typical for transverse coronal loop oscillations, even moderate magnetic twist makes the loop boundary $\sigma$-stable.

In Section \ref{subsec:loop}, we used our model to perform a local stability analysis of the sections of the loop boundary where the amplitudes of the shear flows are the greatest (see Figures \ref{fig:tube1} and \ref{fig:tube_interface}). The local analysis is only valid for perturbations with the azimuthal wavelength much smaller than the radius of the loop cross-section $R$, and the axial wavelength much smaller than the loop length $L$\/. In accordance with these latter assumptions, we took $k_x = m/R$ and $k_z = \pi n/L$, where $k_x$ is the component of the wave vector in the azimuthal direction, and $k_z$ is the component of the wave vector in the axial direction, and $|m|$ and $n$ are positive integer numbers. We note that, while $n$ is positive, $m$ can be either positive or negative. We found that the nature of solutions is changed by this new definition of the parameters. While, previously, all solutions were unstable regardless of the background parameters, the discretisation of the parameter space has introduced the possibility that unstable solutions exist only for sufficiently large values of $|m|$.

{It is worth noting that our study does not include the effects of strong shear induced by resonant absorption, which may be significant in the generation of the KHI, as suggested by \cite{Antolin2014}.
The numerical studies by \cite{Howson2017b} and \cite{Terradas2018} showed that the presence of the transitional layer leads to an increase in the instability growth time.
This suggests that the main driver of the KH instability is the shear motion at the magnetic tube boundary due to the transverse oscillation, as opposed to the shearing caused by resonant absorption.}

{
Our model may be expanded such that more accurate quantitative results about transverse loop oscillations are obtained.
A transitional layer, where the oscillating velocity continuously varies from one side to the other, may be included.
A further extension may consider a continuous variation of density from one side to the other, such that the effects of resonant absorption are also considered.
Both of these generalizations are likely to be mathematically complicated.
}

{
A different possible application of the present model relates to prominence oscillations \citep{Arregui2012}.
Assuming that the magnetic field has the same magnitude inside and outside the structure, for a typical density contrast of $r=100$, Equation \eqref{eq16} yields that $K \ll 1$, for $\theta \neq 0$.
This suggests that, unless the magnetic fields inside and outside the prominence are perfectly aligned, the growth time of perturbations is very small.
}

{Finally, we make the following comment.
Usually it is written in papers dealing with the numerical study of the KH instability of oscillating coronal magnetic loops that this instability occurs in the nonlinear regime.
However, in our paper the background state is given by the linear solution describing the kink oscillation.
The stability analysis is also based on the use of the linear MHD.
This clearly shows that the KH instability of oscillating coronal loops is not related to the nonlinearity at all.}   

\acknowledgments
\noindent
{\bf Acknowledgments:}
MB, MSR and RE are grateful to the Science and Technology Facilities Council (STFC, grant number ST/M000826/1) UK and the Royal Society (UK) for the support received.
TVD was supported by GOA-2015-014 (KU~Leuven). This work was based on discussions at the ISSI (Bern, Switzerland, March 2017). This project has also received funding from the European Research Council (ERC) under the European Union's Horizon 2020 research and innovation programme (grant agreement No 724326).

\appendix

%=============================================================
\section{The Maximum Growth Rate}
\label{sec:appendix}
%=============================================================

As we have already stated before, the characteristic exponent, $\mu$, is determined by the equation
\begin{equation}
\label{eq:A1}
\cosh(\pi \mu) = \bar\eta(\pi),
\end{equation}
where $\bar\eta(\tau)$ is the solution to the initial value problem to Equation~\eqref{eq10} with
\begin{equation}
\label{eq:A2}
\bar\eta = 1, \quad \frac{\mathrm{d} \bar\eta}{\mathrm{d} t} = 0 \quad 
\mbox{at} \;\; \tau = 0,
\end{equation}
\citep{Abramowitz1965}.
When a perturbation is unstable, its growth rate is given by $\gamma = \Re(\mu)$.
In the context of the $\sigma$-stability analysis, we assume that the growth time of the instability is much larger than the oscillation period. In terms of dimensionless quantities, this condition is written as $\gamma \ll 1$. The numerical investigation shows that this condition is only satisfied for all values of $q$ when $K \gg 1$. In accordance with this, we introduce the small parameter $\epsilon = 1 / K$\/. Figure~\ref{fig:stability_diagram} shows that $a$ is close to $j^2$ on parts of the line $a = Kq$ corresponding to unstable perturbations when $K \gg 1$, where $j = 1,2,\dots$ We obtain $a = j^2$ taking $q = j^2\epsilon$\/, which implies that $q = {\cal O}(\epsilon)$.

First we study the case with $j = 1$. Using the expansion valid for small $q$ \citep{Abramowitz1965},
\begin{equation}
\label{eq:A3}
a_1(q) = 1 + q + \mathcal{O}(q^2), \quad b_1(q) = 1 - q + \mathcal{O}(q^2),
\end{equation}
we obtain that the line $a = Kq$ in Figure~\ref{fig:stability_diagram} intersects the curves $a = b_1(q)$ and $a = a_1(q)$ at $q \approx \epsilon - {\cal O}(\epsilon^2)$ and $q \approx \epsilon + {\cal O}(\epsilon^2)$, respectively. Then $q = \epsilon + \bar q\epsilon^2$ on the part of the curve $a = Kq$ between the intersection points, where $\bar q$ is a free parameter varying from approximately $-1$ to approximately 1. It follows that $q = \epsilon + \bar q \epsilon^2$ on the line $a = K q$ between the intersection points, where $\bar q$ is a free parameter. The equation of the curve $a = Kq$ is now rewritten as $a = 1 + \bar q\epsilon$\/, and Equation \eqref{eq10} becomes
\begin{equation}
\frac{\mathrm{d}^2 \eta}{\mathrm{d}\tau^2} + [1 + \bar q\epsilon
- 2(\epsilon + \bar q\epsilon^2)(\cos(2 \tau)] \eta = 0.
\label{eq:A4} 
\end{equation}
To calculate the increment we need to find the solution $\bar\eta(\tau)$ to this equation satisfying the initial conditions Equation~\eqref{eq:A2}. To do this we use the regular perturbation method with
\begin{equation}
\bar\eta = \bar\eta^{(0)} + \bar\eta^{(1)} + \bar\eta^{(2)} + \dots.
\label{eq:A5} 
\end{equation}
Substituting Equation~\eqref{eq:A5} into Equations~\eqref{eq:A2} and \eqref{eq:A4}, and collecting the terms of the order of unity, we obtain
\begin{equation}
\frac{\mathrm{d}^2\bar\eta^{(0)}}{\mathrm{d}\tau^2} + \bar\eta^{(0)} = 0,
\label{eq:A6} 
\end{equation}
and the associated initial conditions
\begin{equation}
\bar\eta^{(0)} = 1, \quad \frac{\mathrm{d} \bar\eta^{(0)}}{\mathrm{d} \tau} = 0
\quad \mbox{at} \;\; \tau = 0.
\label{eq:A7} 
\end{equation}
The solution to this initial value problem is
\begin{equation}
\bar\eta^{(0)} = \cos\tau.
\label{eq:A8}
\end{equation}
Collecting term of the order of $\epsilon$ yields
\begin{equation}
\frac{\mathrm{d}^2\bar\eta^{(1)}}{\mathrm{d}\tau^2} + \bar\eta^{(1)} = 
   [2\cos(2\tau) - \bar q]\cos\tau ,
\label{eq:A9} 
\end{equation}
\begin{equation}
\bar\eta^{(1)} = 0, \quad \frac{\mathrm{d}\bar\eta^{(1)}}{\mathrm{d}\tau} = 0 
   \quad \mbox{at} \;\; \tau = 0.
\label{eq:A10}
\end{equation}
After straightforward calculation we obtain
\begin{equation}
\bar\eta^{(1)} = \frac{1 - \bar q}2\tau\sin\tau - \frac18\cos(3\tau) + \frac18\cos\tau.
\label{eq:A11}
\end{equation}
Finally we collect terms of the order of $\epsilon^2$ to obtain
\begin{equation}
\frac{\mathrm{d}^2\bar\eta^{(2)}}{\mathrm{d}\tau^2} + \bar\eta^{(2)} = 
   [2\cos(2\tau) - \bar q]\eta_1^{(1)} + 2\bar q\cos(2\tau)\cos\tau,
\label{eq:A12} 
\end{equation}
\begin{equation}
\bar\eta^{(2)} = 0, \quad \frac{\mathrm{d}\bar\eta^{(2)}}{\mathrm{d}\tau} = 0 
   \quad \mbox{at} \;\; \tau = 0.
\label{eq:A13}
\end{equation}
The solution to this initial value problem is given by
\begin{eqnarray}
\bar\eta^{(2)} &=& \frac{1 - \bar q^2}8\tau^2\cos\tau + 
   \frac{2\bar q^2 + 7\bar q -2}{16}\tau\sin\tau - 
   \frac{1 - \bar q}{16}\tau\sin(3\tau) \nonumber\\ 
&+& \frac{\cos(5\tau)}{192} - \frac{2 + 3\bar q}{32}\cos(3\tau) + 
   \frac{11 + 18\bar q}{192}\cos\tau.
\label{eq:A14}
\end{eqnarray}
Using Equations~\eqref{eq:A8}, \eqref{eq:A11}, and \eqref{eq:A14} we obtain
\begin{equation}
\bar\eta(\pi) = -1 - \frac{1 - \bar q^2}8\pi^2\epsilon^2 + {\cal O}(\epsilon^3).
\label{eq:A15}
\end{equation}
It follows from this equation that 
\begin{equation}
\mu = i \pm \frac\epsilon2\sqrt{1 - \bar q^2} + {\cal O}(\epsilon^2).
\label{eq:A16}
\end{equation}
This result implies that
\begin{equation}
\gamma = \frac\epsilon2\sqrt{1 - \bar q^2} + {\cal O}(\epsilon^2), \quad
   \gamma_m = \frac\epsilon2 ,
\label{eq:A17}
\end{equation}
where $\gamma_m$ is the maximum value of the instability increment when the point $(a,q)$ is on the part of line $a = Kq$ that is between the curves $a = b_1(q)$ and $a = a_1(q)$.

Now we consider the part of line $a = Kq$ that is between the curves $a = b_j(q)$ and $a = a_j(q)$, $j = 2,3,\dots$ For $q \ll 1$ we have $b_1(q) = n^2 + {\cal O}(q^2)$ and $a_1(q) = n^2 + {\cal O}(q^2)$, where $n$ is a natural number \citep{Abramowitz1965}. Since $K = \epsilon^{-1}$\/, it follows that $q = n^2\epsilon(1 + \bar q\epsilon^2)$ and $a = n^2(1 + \bar q\epsilon^2)$, where $\bar q$ is again a free parameter. Substituting these expressions in Eq.~(\ref{eq:A1}) we transform it to
\begin{equation}
\frac{\mathrm{d}^2\eta}{\mathrm{d}\tau^2} + j^2[1 + \bar q\epsilon^2 - 
   2(\epsilon + \bar q\epsilon^3)(\cos(2 \tau)] \eta = 0.
\label{eq:A18} 
\end{equation}
Then we again look for the solution in the form of the expansion given by Eq.~(\ref{eq:A5}). Substituting this expansion in Equations \eqref{eq10} and \eqref{eq:A2}, and collecting terms of the order of unity we obtain
\begin{equation}
\frac{\mathrm{d}^2\bar\eta^{(0)}}{\mathrm{d}\tau^2} + j^2\bar\eta^{(0)} = 0,
\label{eq:A19} 
\end{equation}
\begin{equation}
\bar\eta^{(0)} = 1, \quad \frac{\mathrm{d}\bar\eta^{(0)}}{\mathrm{d}\tau} = 0 
   \quad \mbox{at} \;\; \tau = 0.
\label{eq:A20}
\end{equation}
The solution to this initial value problem is
\begin{equation}
\bar\eta^{(0)} = \cos(j\tau).
\label{eq:A21}
\end{equation}
Collecting terms of the order of $\epsilon$ yields
\begin{equation}
\frac{\mathrm{d}^2\bar\eta^{(1)}}{\mathrm{d}\tau^2} + j^2\bar\eta^{(1)} = 
   2j^2\cos(2\tau)\cos(j\tau) ,
\label{eq:A22} 
\end{equation}
\begin{equation}
\bar\eta^{(1)} = 0, \quad \frac{\mathrm{d}\bar\eta^{(1)}}{\mathrm{d}\tau} = 0 
   \quad \mbox{at} \;\; \tau = 0.
\label{eq:A23}
\end{equation}
After straightforward calculation we obtain
\begin{equation}
\bar\eta^{(1)} = 1 - \frac13\cos(4\tau) - \frac23\cos(2\tau)
\label{eq:A24}
\end{equation}
for $j = 2$, and 
\begin{equation}
\bar\eta^{(1)} = \frac{j^2\cos[(j-2)\tau]}{4(j - 1)} - 
   \frac{j^2\cos[(j+2)\tau]}{4(j + 1)} - \frac{n^2\cos(j\tau)}{2(j^2 - 1)}
\label{eq:A25}
\end{equation}
for $j > 2$. Collecting terms of the order of $\epsilon^2$ we obtain
\begin{equation}
\frac{\mathrm{d}^2\bar\eta^{(2)}}{\mathrm{d}\tau^2} + \bar\eta^{(2)} = 
   2j^2\bar\eta^{(1)}\cos(2\tau) - j^2\bar q\cos(j\tau),
\label{eq:A26} 
\end{equation}
\begin{equation}
\bar\eta^{(2)} = 0, \quad \frac{\mathrm{d}\bar\eta^{(2)}}{\mathrm{d}\tau} = 0 
   \quad \mbox{at} \;\; \tau = 0.
\label{eq:A27}
\end{equation}
The solution to this initial value problem is given by
\begin{equation}
\bar\eta^{(2)} = \left(\frac53 - \bar q\right)\tau\sin(2\tau) + 
   \frac{\cos(6\tau)}{24} + \frac29\cos(4\tau) + 
   \frac{29}{72}\cos(2\tau) - \frac23
\label{eq:A28}
\end{equation}
for $j = 2$, and by
\begin{eqnarray}
\bar\eta^{(2)} &=& \frac j4\left(\frac{j^2}{j^2 - 1}\right)\tau\sin(j\tau) + 
   \frac{j^4\cos[(j+4)\tau]}{32(j+1)(j+2)} + 
   \frac{j^4\cos[(j+2)\tau]}{8(j+1)(j^2 - 1)}\nonumber\\ 
&-& \frac{j^4(j^4 - 3j^2 + 16)\cos(j\tau)}{16(j^2 - 1)^2(j^2 - 4)} -
   \frac{j^4\cos[(j-2)\tau]}{8(j-1)(j^2 - 1)} + \frac{j^4\cos[(j+4)\tau]}{32(j-1)(j-2)}
\label{eq:A29}
\end{eqnarray}
for $j > 2$. Using Eqs.~(\ref{eq:A21}), (\ref{eq:A24}), (\ref{eq:A25}), (\ref{eq:A28}), and (\ref{eq:A29}), we obtain
\begin{equation}
\bar\eta(\pi) = (-1)^n + {\cal O}(\epsilon^3).
\label{eq:A30}
\end{equation}
It follows from this equation that $\mu = {\cal O}(\epsilon^3)$ for even $j$ and $\mu = i + {\cal O}(\epsilon^3)$ for odd $j$\/, and thus $\gamma = {\cal O}(\epsilon^{3/2})$, that is $\gamma \ll \gamma_m$\/. Hence, $\gamma_m = 1/2K$ is the maximum value of the instability increment with respect to $q$ when $K = \epsilon^{-1}$\/. 

\bibliographystyle{aasjournal}
\bibliography{bibrefs}

\end{document}